\documentclass[10pt]{article}
\usepackage[dvips,usenames]{color}
\usepackage{multicol}
\usepackage{latexsym}
\usepackage{amssymb}
\usepackage{amsmath}
\usepackage{a4wide}
\usepackage{psfig}

\def\doublespace{\def\baselinestretch{1.5}}
\doublespace

\begin{document}

\begin{center}
  {\bf Autocatalytic reaction front in a pulsative periodic flow}\\
  \vspace{0.1in}
  by\\
  M. Leconte, J. Martin, N. Rakotomalala, D. Salin\\
  Laboratoire Fluides Automatique et Syst\`emes Thermiques,\\
  Universit\'es P. et M. Curie and Paris Sud, C.N.R.S. (UMR 7608),\\
  B\^atiment 502, Campus Universitaire, 91405 Orsay Cedex,
  France\\

  {\bf ABSTRACT}\\
\end{center}
Autocatalytic reaction fronts between reacted and unreacted
species may propagate as solitary waves, namely at a constant
front velocity and with a stationary concentration profile,
resulting from a balance between molecular diffusion and chemical
reaction. A velocity field in the supporting medium may affect the
propagation of such fronts through different phenomena:
convection, diffusion enhancement, front shape changes... We
report here on an experimental study and lattice BGK numerical
simulations of the effect of an oscillating flow on the
autocatalytic reaction between iodate and arsenous acid in a
Hele-Shaw cell. In the low frequency range covered by the
experiments, the front behavior is controlled by the flow across
the gap and can be reproduced with 2D numerical simulations. An
analytical solution is also derived under the condition of a weak
flow velocity, and is found to be in reasonable agreement with our
data.

\today

\section*{Introduction}

Interface motion and reaction front propagation occur in a number
of different areas \cite{scott94}, including flame propagation in
combustion \cite{clavin85}, population dynamics \cite{fisher37,
kpp37} and atmospheric chemistry (ozone hole). An autocatalytic
reaction front between two reacting species propagates as a
solitary wave, that is, at a constant front velocity and with a
stationary front profile \cite{epik55, hanna82}. These issues were
addressed earlier on, but only a few cases are understood, such as
the pioneering works of Fisher \cite{fisher37} and
Kolmogorov-Petrovskii-Piskunov \cite{kpp37} on a
reaction-diffusion equation with a second-order kinetics
\cite{scott94, zeldovitch38, ebert00}. Although the effect of an
underlying flow on a flame propagation has been extensively
analyzed \cite{clavin85, zeldovitch38}, the advective effect on
the behavior of an autocatalytic front has been addressed only
recently \cite{papanicolaou91,audoly00,abel01,nolen03}. In this
case, the evolution of the concentration of each chemical species
is given by the Advection-Diffusion-Reaction (ADR) equation:
\begin{eqnarray}
  \frac{\partial C}{\partial t}+
  \overrightarrow{U}.\overrightarrow{\triangledown} C =
  D_m\triangle C + \alpha f(C)
  \label{adrevect}
\end{eqnarray}
where $C$ is the normalized concentration of the (autocatalytic)
reactant, $\overrightarrow{U}$ is the flow velocity, $D_m$ is the
molecular diffusion coefficient and $\alpha$ is the reaction rate.\\
In the absence of flow ($\overrightarrow{U}=\overrightarrow{0}$),
the balance between diffusion and reaction leads to a solitary
wave of constant velocity $V_\chi$ and width $l_\chi$. For the
autocatalytic Iodate-Arsenous Acid (IAA) reaction studied here,
the kinetics is of the third order \cite{scott94},
$f(C)=C^2(1-C)$, and the following $1D$ solution of equation
(\ref{adrevect}) is obtained \cite{hanna82,bockmann00}:
\begin{equation}
  C(z,t)=\frac{1}{1+e^{(z-V_\chi t)/l_\chi}}\;\;,\;\;
  l_\chi=\sqrt{\frac{2D_m}{\alpha}}\;\;,\;\;
  V_\chi=\sqrt{\frac{\alpha D_m}{2}}
  \label{co}
\end{equation}
where $z$ is the direction of the front propagation.\\

For a reaction propagating along the direction of a unidirectional
stationary flow, $\overrightarrow{U}$, two regimes have been
described, depending on the ratio  $\eta=b/2l_\chi$, where $b$ is
the typical length scale transversely  to the flow direction
\cite{edwards02,leconte03,leconte04}. In the eikonal regime, $\eta
\gg 1$, the front propagates as a planar wave, at a velocity given
by the sum of $V_\chi$ and of the algebraic maximum of the flow
velocity (projected onto the direction of
$\overrightarrow{V_\chi}$), and takes the according stationary
form. In the mixing regime, $\eta \ll 1$, the interplay between
diffusion and advection enhances the mixing of the chemical
species and leads to an overall macroscopic diffusion known as
Taylor hydrodynamic dispersion \cite{taylor54}. As a result, the
front moves faster. However, it is still described by equation
(\ref{co}), in which the molecular diffusion coefficient $D_m$ has
to be replaced by its effective macroscopic counterpart.

The main idea of the present paper is to address the effect of an
unsteady flow  on the front propagation. We measure,
experimentally and numerically, the velocity and width of a
chemical front submitted to a time periodic flow, of period
$T=1/f=2\pi/\omega$. The question of the relevant time scale, to
which the time scale of the flow, $T$, has to be compared, is
discussed.

We extend the theoretical work by Nolen and Xin {\it et al.}
\cite{nolen03}, who derived the time-averaged chemical front
velocity in an oscillating flow, by analyzing the temporal
evolution of the front velocity. We note that in the tracer case
(without reaction), Chatwin \cite{chatwin75} and Smith
\cite{smith82} showed, using a Taylor-like approach
\cite{taylor54}, that a pulsating flow results in an effective
time dependent diffusion coefficient, the time-average of which is
larger than the molecular diffusion coefficient \cite{watson83}.\\

In this paper, we study, experimentally and numerically, a
third-order autocatalytic Iodate-Arsenous Acid (IAA) reaction
submitted to a pulsative flow. In section $1$, we present the
experimental set-up and the measurements  obtained using a large
set of frequencies and amplitudes of oscillations. In section $2$,
we compare the experimental results with $2D$ numerical
simulations and we investigate a wider range of parameters with
additional simulations. In the last section, we extend the
theoretical result by Nolen and Xin \cite{nolen03} to derive the
temporal variations of the front velocity.

\section*{Experimental set-up and data}

\begin{figure}[!h]
  \begin{center}
    \begin{minipage}{100mm}
      \psfig{file=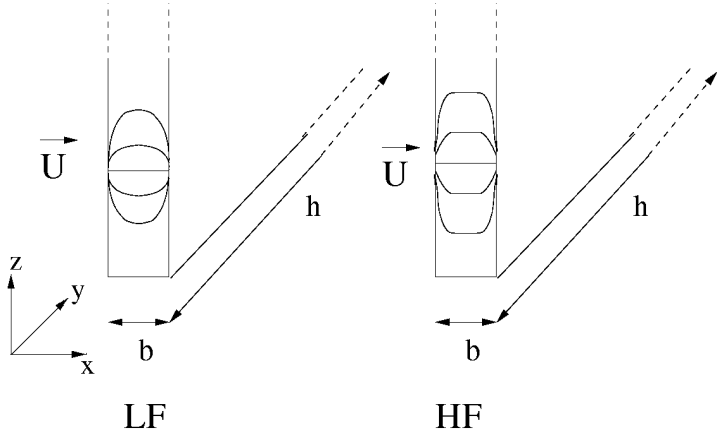,width=100mm,angle=-90}
    \end{minipage}
  \end{center}
  \caption{\small Sketch of the velocity profile in the gap of a Hele-Shaw cell for low
    frequency (LF, left) and high frequency (HF, right).}
  \label{fig_oscill}
\end{figure}

We use the third-order autocatalytic Iodate-Arsenous Acid (IAA)
reaction. The reaction front is detected with starch, leading to a
dark blue signature of the transient iodine as the reaction occurs
\cite{scott94,hanna82,bockmann00}. In the absence of flow, a
reaction front travels like a solitary wave, with a constant
velocity $V_\chi^{exp}\sim 20\; \mu m/s$ and with a stationary
concentration profile of width $l_\chi^{exp}\sim 100\; \mu m$. We
study the front propagation in a  Hele-Shaw (HS) cell of
cross-section $b\times h=0.4\times 8\; mm^2$ (along $x$ and $y$
directions, respectively). The unidirectional (along $z$
direction) oscillating flow is imposed at the bottom of a vertical
HS cell, from a reservoir filled with unreacted species. This
revervoir is closed with a thin  elastic membrane, pressed in its
middle by a rigid rod fixed at the center of a loudspeaker.
Consequently, a displacement of a given volume of liquid in the
reservoir induces a displacement of the same volume of liquid in
the HS cell. The $y-z$ plane of the HS cell is enlightened from
behind and recorded with a CCD camera.  The amplitude $A$ and the
pulsation $\omega=2\pi f$ of the oscillating flow are imposed by
the controlled sine tension applied to the loudspeaker, and
measured in situ, from the recorded displacement of the air/liquid
interface at the top of the partially filled HS cell. This
displacement follows the expected $A\sin{(\omega t)}$ time
dependence. Due to the constraint of our device, the imposed
amplitude and frequency of the flow displacement are in the ranges
$A\in [0.07\; mm,1.7\; mm]$ and $f\in [0.01\; Hz, 0.2\; Hz]$ and
the maximum velocity of the flow in the cell is roughly
$U_M=A\omega$. The oscillating flow field in the HS cell is of the
form (see Appendix for the full expression):
\begin{equation}
  U(x,y,t)=U_{M} \Re e [f(x,y)]\sin{(\omega t)}
  \label{soleqfinale}
\end{equation}
The shape of the velocity profile depends drastically on the
viscous penetration length $l_\nu=\sqrt{\nu/\omega}$
\cite{landau89}. If $l_\nu$ is large compared to the cell
thickness $b$ (low frequency), the flow variations are slow enough
for the steady state to be established. The resulting "oscillating
stationary" velocity profile is parabolic in the gap and flat
along the width $h$ of the cell except in the vicinity of the side
walls, in a layer of thickness $b$ \cite{gondret97}. Conversely,
for $l_\nu<<b$ (high frequency), the fluid has not enough time to
feel the effects of the solid boundaries and the velocity profile
is flat over the whole cross-section $b\times h$, except in the
vicinity of each wall, in a layer of thickness $l_\nu$. Figure
\ref{fig_oscill} is a sketch of such an effect. For our dilute
aqueous solutions of viscosity $\nu \approx 10^{-6}\; m^2.s^{-1}$
and in our frequency range, the penetration length $l_\nu$, which
ranges between $0.8\; mm$ and $4\; mm$, is larger than the cell
thickness $b=0.4\; mm$.

Hence, in most of our experiments, the "stationary velocity
profile" is instantaneously reached, parabolic Poiseuille-like
across the gap $b$, and almost invariant along the $y$ direction
(except in a layer of thickness $l_\nu$  close to the boundaries).\\

\begin{figure}[!h]
  \begin{center}
    \begin{minipage}{100mm}
      \psfig{file=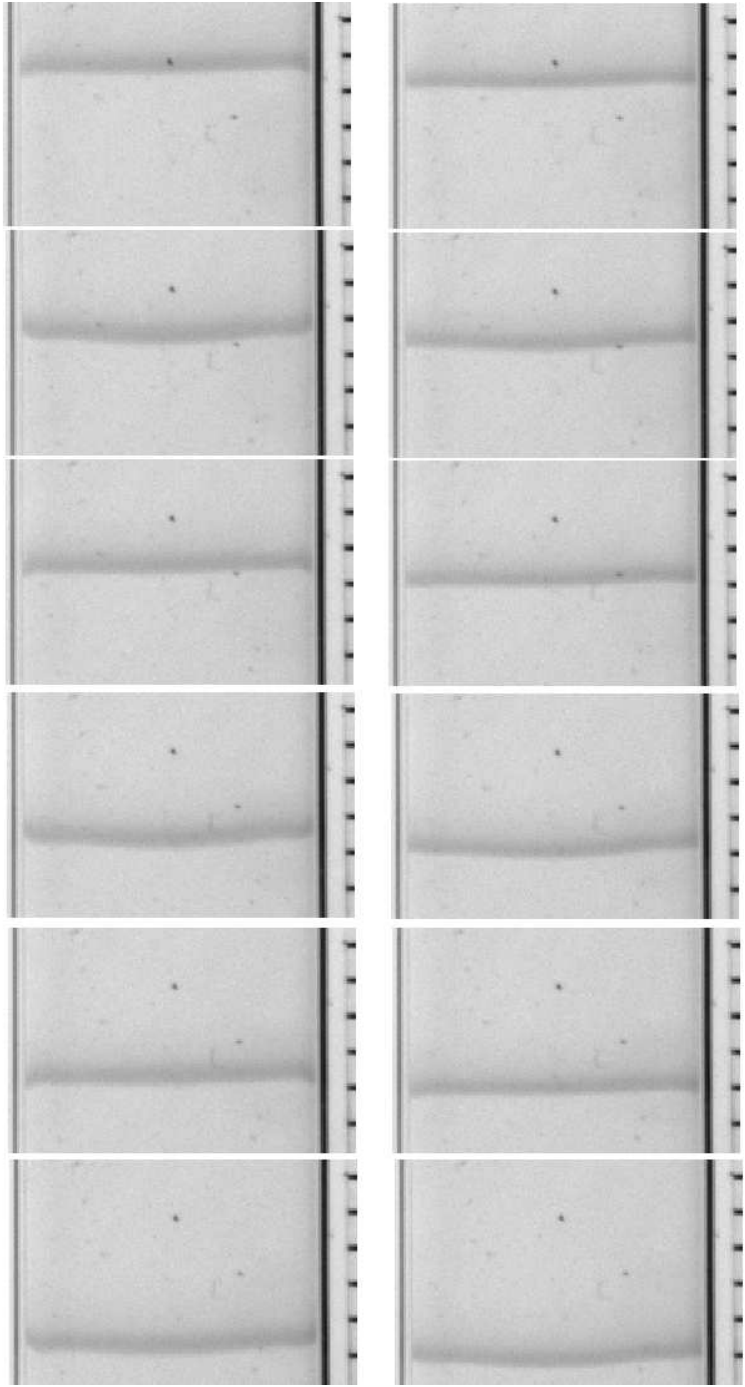,width=100mm,angle=0}
    \end{minipage}
  \end{center}
  \caption{\small Time evolution of a chemical front in a pulsative flow field of
    amplitude $A=0.55\; mm$ and period $T=50\; s$. Time increases
    from left to right and from top to bottom. Two images are separated  by $T/4$ time intervals.
    The distance between two dashes is $1\; mm$.}
  \label{ext_larg}
\end{figure}

Figure \ref{ext_larg} displays snapshots of a typical experiment:
We do observe a front slightly deformed,  propagating up and down
(oscillating), with a downward averaged displacement, from the
burnt product of the reaction to the fresh reactant.

\begin{figure}[!h]
  \begin{minipage}{160mm}
      \psfig{file=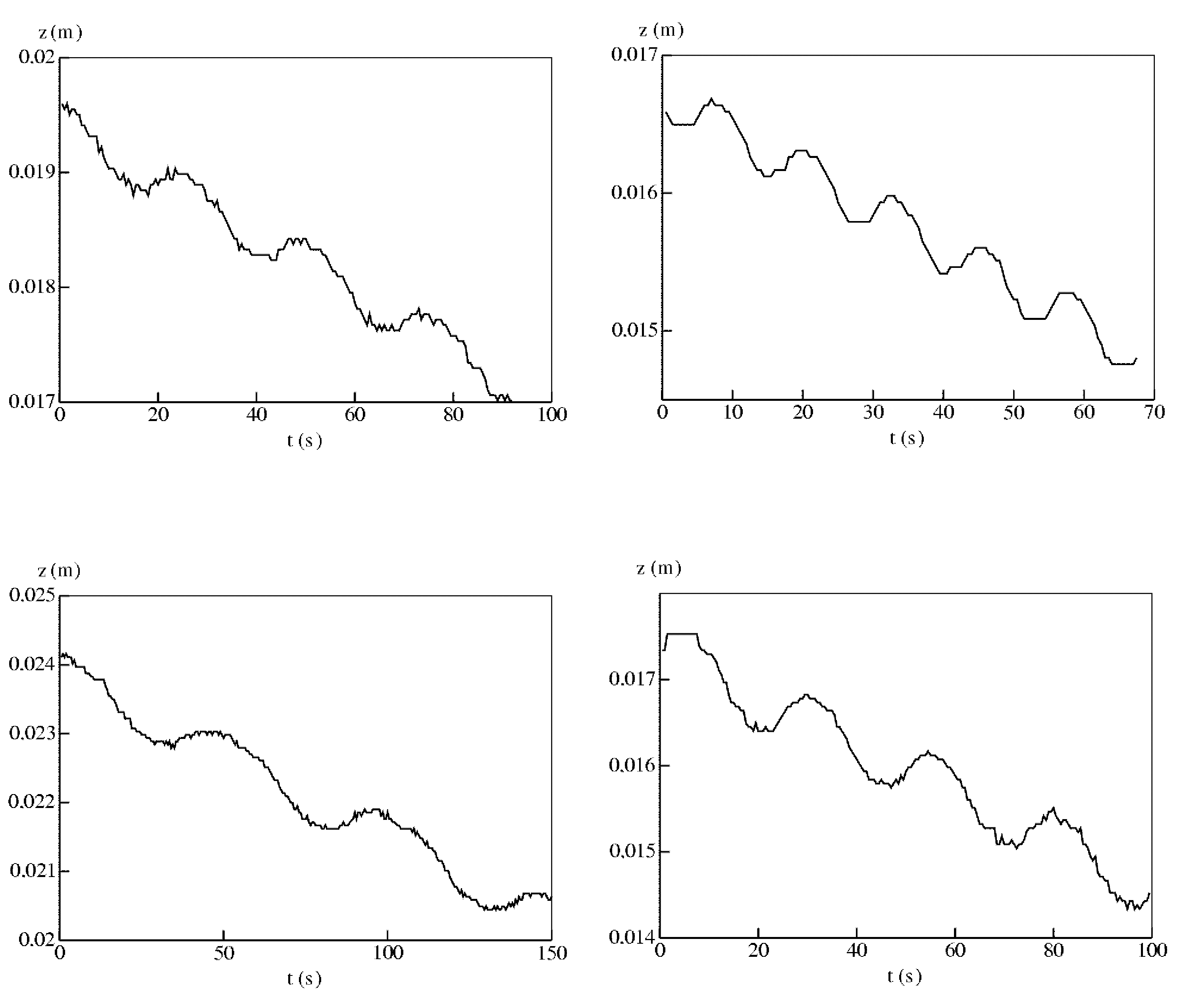,width=160mm,angle=-90}
  \end{minipage}
  \caption{\small Front displacement for different couples $(A,f)$
    ($A$ in $mm$ and $f$ in $Hz$).
    Top : $(0.28,0.04)$, $(0.55,0.02)$.
    Bottom : $(0.28,0.08)$, $(0.55,0.04)$.}
  \label{depla_exp}
\end{figure}

From this movie, the front is tracked and its location is plotted
as a function of time. The so-obtained figure \ref{depla_exp}
clearly shows the oscillation of the front position at roughly the
imposed frequency and an overall drift of the front.

\begin{figure}[!h]
  \begin{center}
    \begin{minipage}{110mm}
      \psfig{file=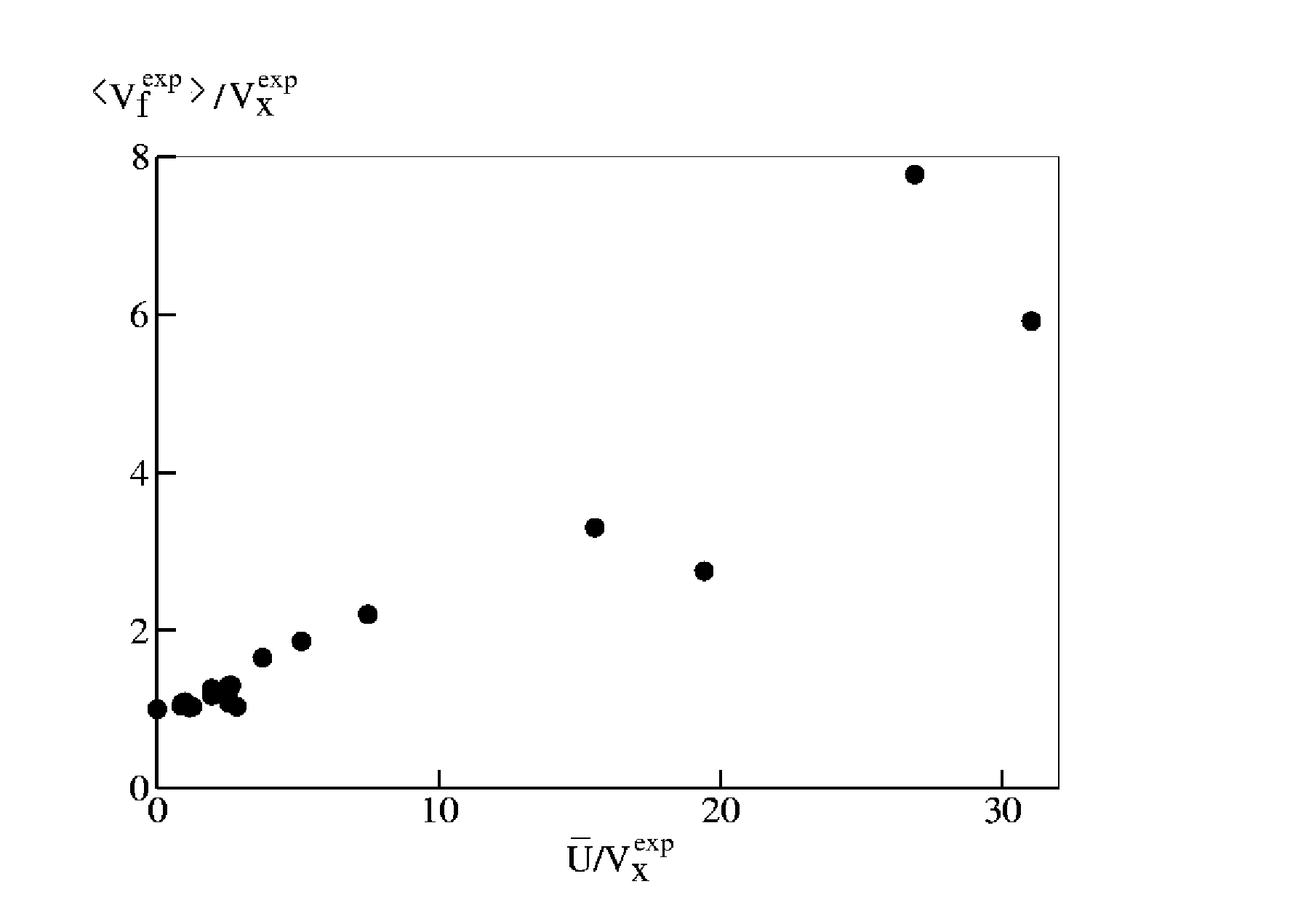,width=110mm,angle=-90}
    \end{minipage}
  \end{center}
  \caption{\small Normalized
     drift velocity of the chemical front
     $\langle V_f^{exp}\rangle/V_\chi^{exp}$ versus
    the normalized flow intensity $\overline{U}/V_\chi^{exp}$.}
  \label{vit_moy_exp}
\end{figure}

The measurement of this drift in time leads to the time-averaged
front velocity $\langle V_f^{exp}\rangle$. Figure
\ref{vit_moy_exp} displays $\langle V_f^{exp}\rangle$, normalized
by $V_{\chi}^{exp}$, versus the amplitude of the time-periodic
flow field $\overline{U}=2U_M/3=2A\omega /3$ (where $2/3$ is the
ratio of the gap-averaged velocity to the maximum one of a $2D$
gap Poiseuille profile), also normalized by $V_{\chi}^{exp}$. The
increase of $\langle V_f^{exp}\rangle$ with $\overline{U}$ is
almost linear, with a slope slightly larger than $0.1$. This
demonstrates that the propagation velocity of a reaction front can
be enhanced by a null in average, laminar flow. Moreover, as the
mean advection in this time-periodic flow is zero, this effect
comes clearly from some non-linear interplay. It could be
attributed to the enhancement of the mixing due to the presence of
the flow.

As mentioned above, it is seen from the instantaneous velocity
curve ($V_f^{exp}(t)$, figure \ref{depla_exp}), that the front
velocity oscillates at the frequency of the flow. However, due to
the experimental noise, it is difficult to obtain further
information from this curve.

We also noticed from our observation of the experimental movies
that the width of the colored front, $L(t)$, was likely to
oscillate at a frequency twice that of the excitation, which,
unfortunately, is not obvious on the static pictures (figure 2).
We note that this feature could support the description of the
front thickness in the framework of an effective diffusion of
coefficient $D$, as the latter is expected to be insensitive to
the flow direction and to depend only on the flow intensity as $D
\propto U^{2}$ \cite{leconte04} (which here oscillates at $2f$).

\begin{figure}[!h]
  \begin{center}
    \begin{minipage}{120mm}
      \psfig{file=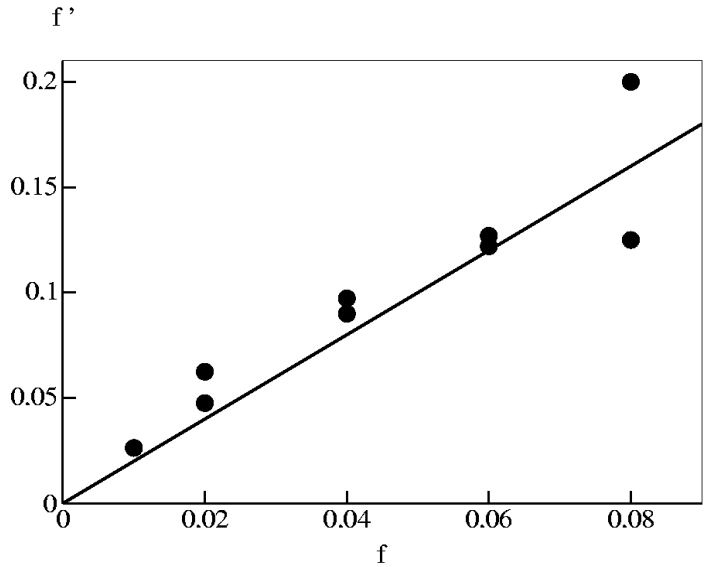,width=120mm,angle=-90}
    \end{minipage}
  \end{center}
  \caption{\small Experimental oscillation frequency $f'$ (Hz) of the front
    width versus the imposed oscillation frequency $f$ of the flow (Hz).
    $\bullet$: Experimental values obtained with the micro-Doppler algorithm,
     \bf{---} : $f'= 2f$.}
  \label{mic_dop}
\end{figure}

To test this empirical observation, we measured the width $L(t)$
of the dark blue ribbon. As this ribbon corresponds to the
presence of the transient iodine, $L(t)$ is a qualitative measure
of the chemical front width, but gives however the right time
behavior. A classical Fourier analysis of $L(t)$ was tried, but,
due to the large amount of noise, it did not provide any reliable
frequency dependence. Therefore, we used the more sensitive
micro-Doppler method (see \cite{millnert95,levyleduc04} and the
references therein) which analyzes an instantaneous signal
frequency. The so-obtained oscillation frequencies $f'$ of the
width $L(t)$ versus the imposed ones $f$, are displayed in figure
\ref{mic_dop}: They collapse onto the straight line $f'=2f$, which
supports the contention that the front width oscillates at twice
the frequency of the flow.

In order to have further insight into the instantaneous features
of the propagating front, we used numerical simulations, which are
less noisy and give access to the behavior in the gap of the cell.

\section*{2-D Numerical simulations}

Assuming a third-order autocatalytic reaction kinetics for the IAA
reaction \cite{scott94,hanna82,bockmann00} and a unidirectional
flow $U(x,y,t)$ in the $z$ direction, equation (\ref{adrevect})
reads:

\begin{eqnarray}
  \frac{\partial C}{\partial t}+U \frac{\partial C}{\partial z}=
  D_m\left (\frac{\partial^2 C}{\partial x^2} +
  \frac{\partial^2 C}{\partial z^2} \right )+\alpha C^2(1-C)
  \label{adr_tempo}
\end{eqnarray}
where $C$ is the concentration of the (autocatalytic) reactant
iodide, normalized by the initial concentration of iodate,
 $D_m$ is the molecular diffusion coefficient and
$\alpha$ is the kinetic rate coefficient of the reaction. The
solution of equation (\ref{adr_tempo}) is obtained using a lattice
Bhatnagar-Gross-Krook (BGK) method, shown to be efficient in
similar contexts
\cite{leconte03,gondret97,martin02,rakotomalala97}. The full $3D$
periodic velocity field $U(x,y,t)$, in a Hele-Shaw cell,
 has been derived analytically in the Appendix
(equation (\ref{soleq2})). As mentioned above, the oscillating
flow field does not vary much along the $y$ direction (except in a
boundary layer of the order of the gap thickness $b$), and the
velocity field, away from the side walls, is basically a $2D$ flow
field, $U(x,t)$, given by:

\begin{equation}
  U(x,t)=U_M
  \Re e \left [\left (1-\frac{\cos{(kx)}}{\cos{(kb/2)}}\right )e^{i\omega t}
    \right ]
  \label{champvit2d}
\end{equation}
where  $k=\sqrt{\frac{i\omega}{\nu}}$ is a complex wave number.
Note that for small frequency, $\omega \ll \nu/b^2$ (i.e. $b\ll
l_{\nu}$), equation (\ref{champvit2d}) reduces to an oscillating
Poiseuille flow: $U(x,t)\approx U_M(1-4x^2/b^2)\sin{(\omega t)}$.
The analytic flow field (equation (\ref{champvit2d})) is used in
equation (\ref{adr_tempo}) for the $2D$ simulation of the ADR
equation by the lattice BGK method.

\begin{figure}[!h]
  \begin{center}
    \begin{minipage}{130mm}
      \psfig{file=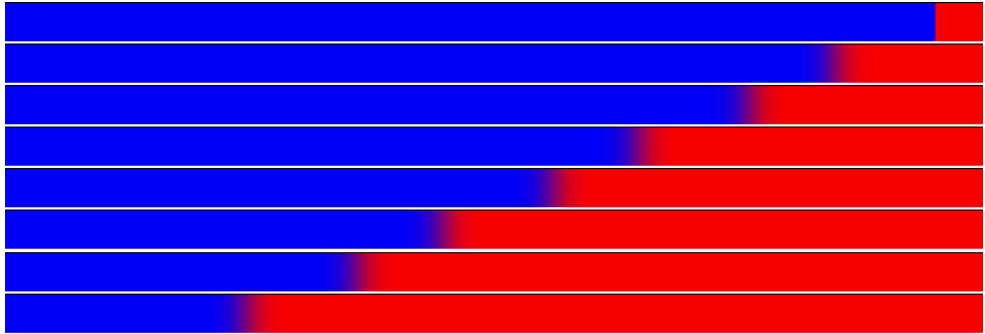,width=130mm,angle=0}
    \end{minipage}
  \end{center}
  \caption{\small Numerical simulation of the front displacement (obtained with $A=10$
    and $f=2.5\times 10^{-6}$). The product of the reaction is in dark
    and the reactant is in grey. From top to bottom, time increases
    by $200000$ time steps ($=1/2$ period). The lattice dimensions
    are $40\times 4000$ (note that the aspect ratio of the pictures is not $1$).
  }
  \label{depla_num}
\end{figure}

In order to compare the results of the numerical simulations with
the experiments, we used the same non dimensional quantities,
namely $b/l_\chi^{exp}$ ($=4$), $U_{M}/V_{\chi}^{exp}$ and the
Schmidt number $Sc=\nu/D_m$ ($=500$) which compares the viscous
and mass diffusivities. The simulations were performed on a
lattice of length $N_z$, ranging between $2000$ and $6000$ nodes,
and of constant width $N_x=40$ nodes during $2\times 10^5$ to
$4\times 10^6$ iterations. The above experimental value of
$b/l_\chi^{exp}$ gives a numerical chemical length
$l_\chi=10=\sqrt{\frac{2D_m}{\alpha}}$. We chose $D_m=5.10^{-3}$
and $\alpha =10^{-4}$, which sets the front velocity
$V_\chi=\sqrt{\frac{\alpha D_m}{2}}=5.10^{-4}$ and the kinematic
viscosity $\nu=D_m Sc=2.5$. The varying parameters in the
simulations are the amplitude $A$ and the frequency $f$ of the
imposed oscillating flow field. A typical movie of a numerical
simulation is displayed in figure \ref{depla_num}.

\begin{figure}[!h]
  \begin{center}
    \begin{minipage}{140mm}
      \psfig{file=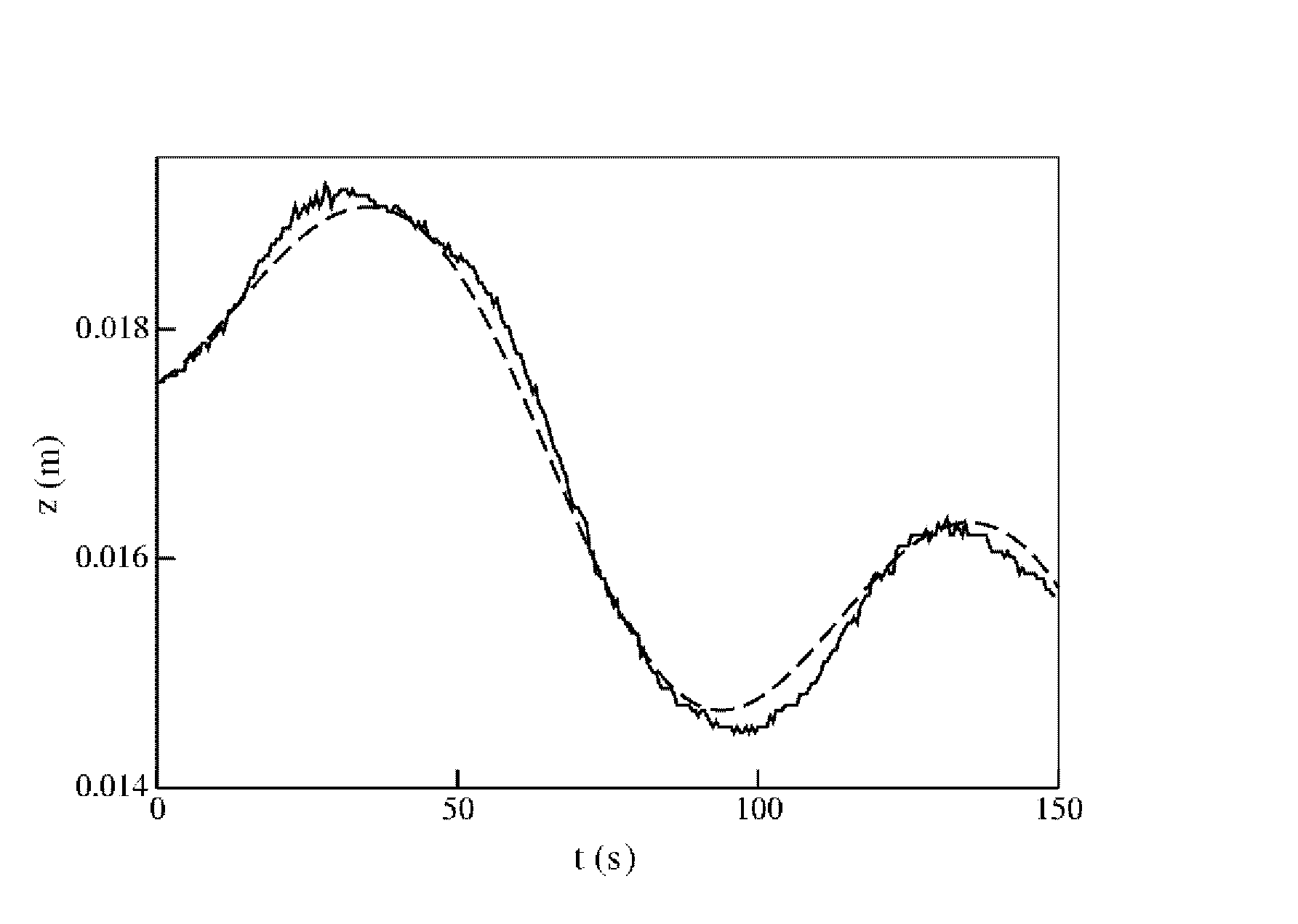,width=70mm,angle=-90}
      \psfig{file=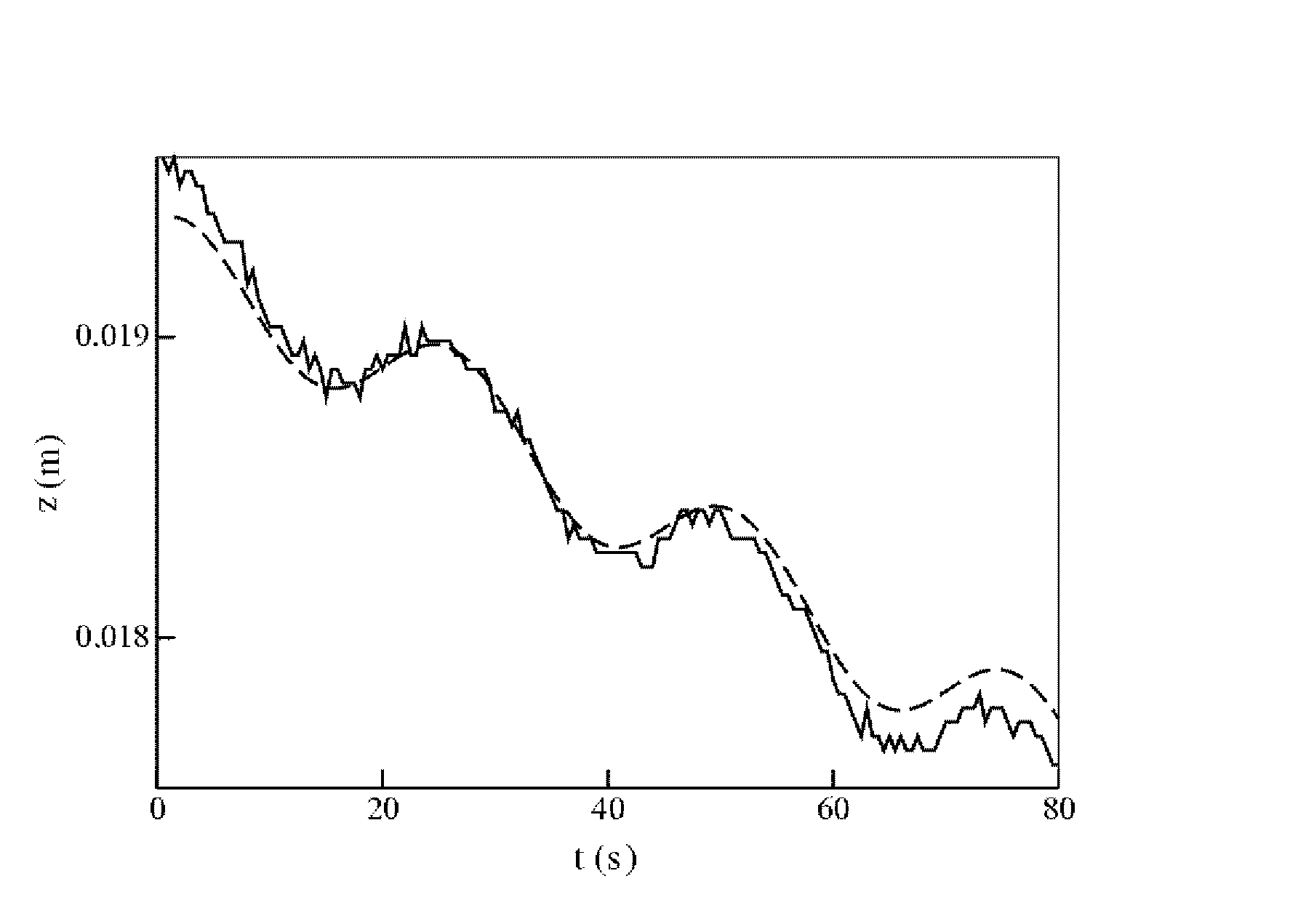,width=70mm,angle=-90}
    \end{minipage}
    \begin{minipage}{140mm}
      \psfig{file=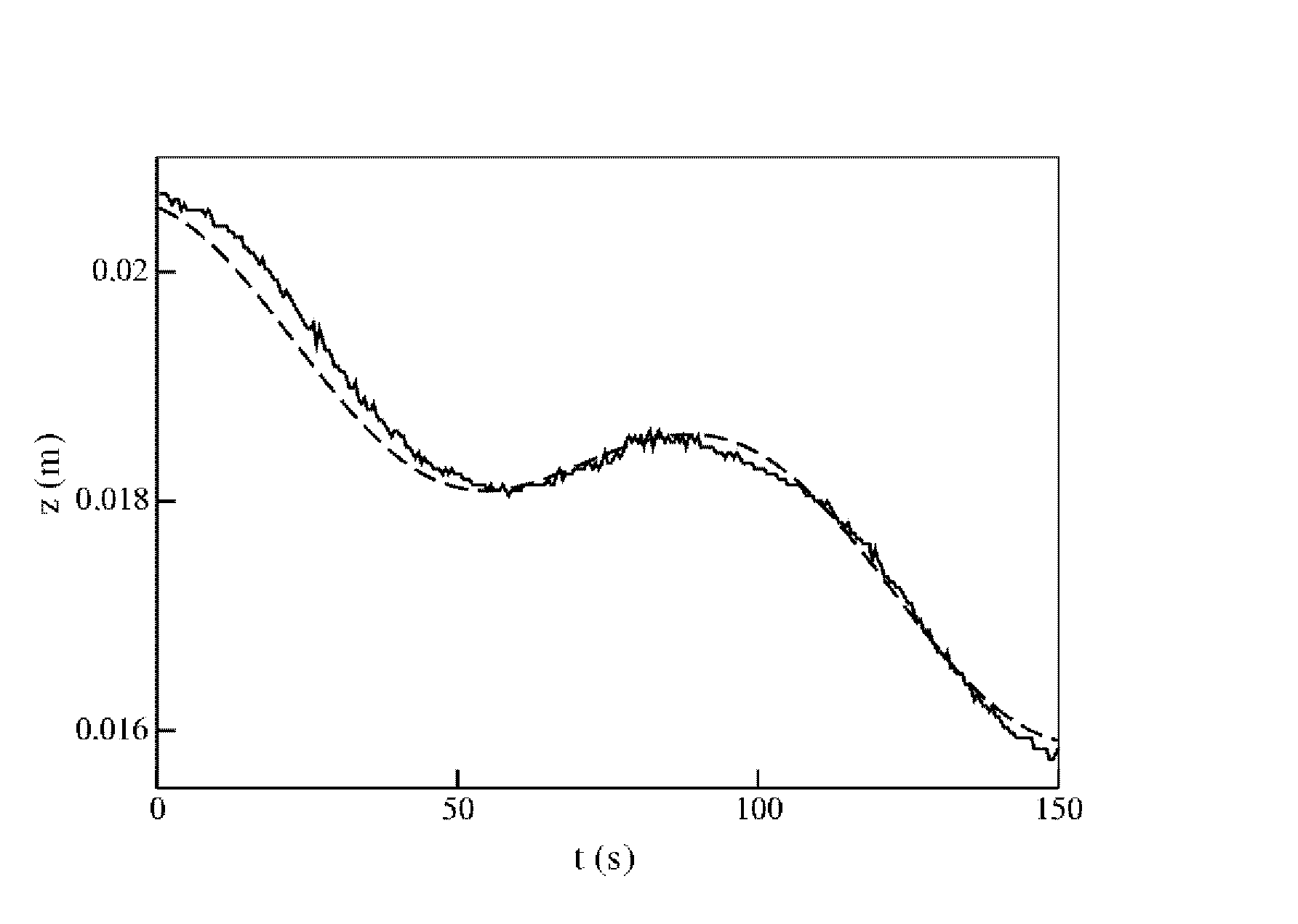,width=70mm,angle=-90}
      \psfig{file=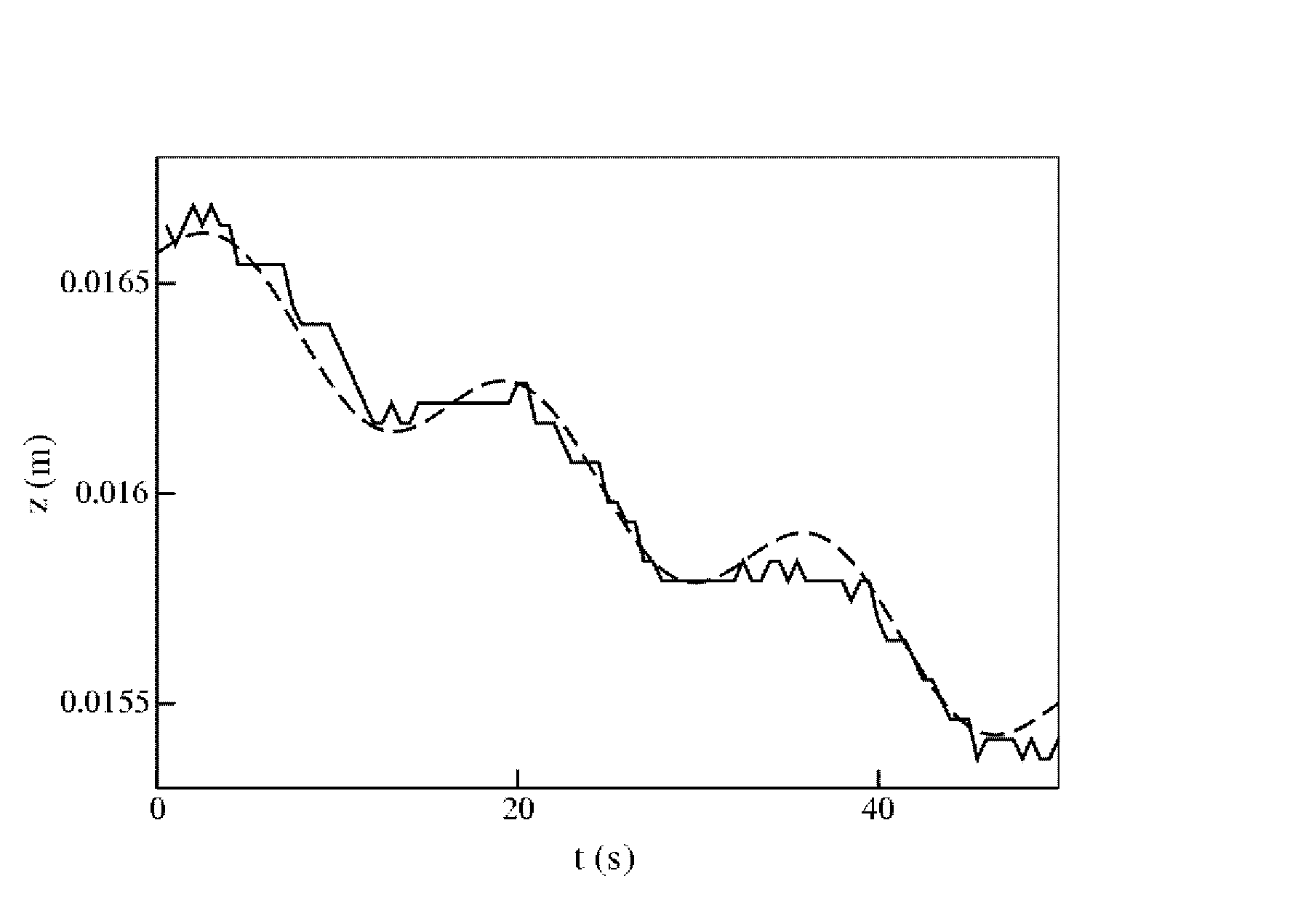,width=70mm,angle=-90}
    \end{minipage}
    \begin{minipage}{140mm}
      \hspace{-1mm} \psfig{file=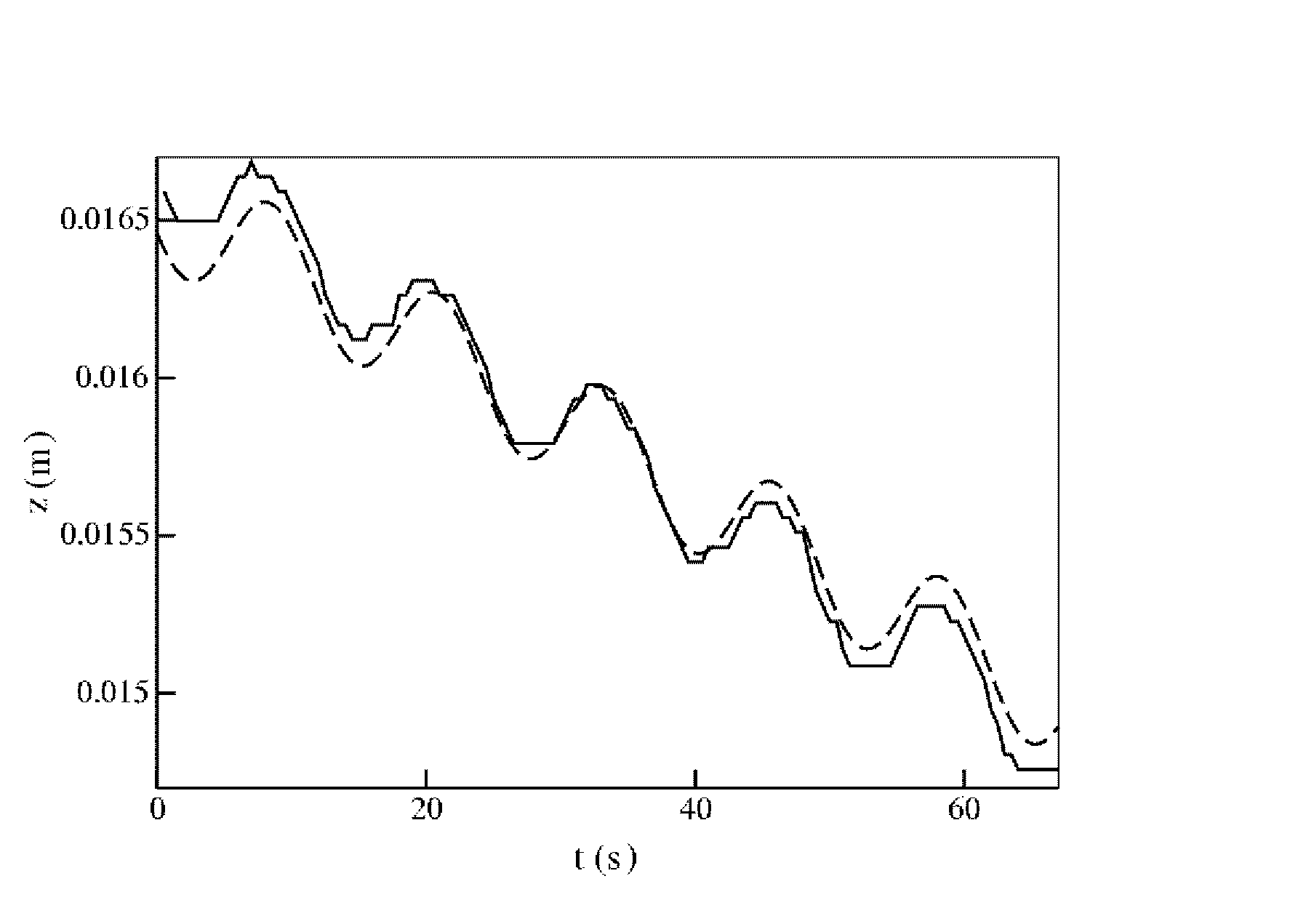,width=70mm,angle=-90}
      \psfig{file=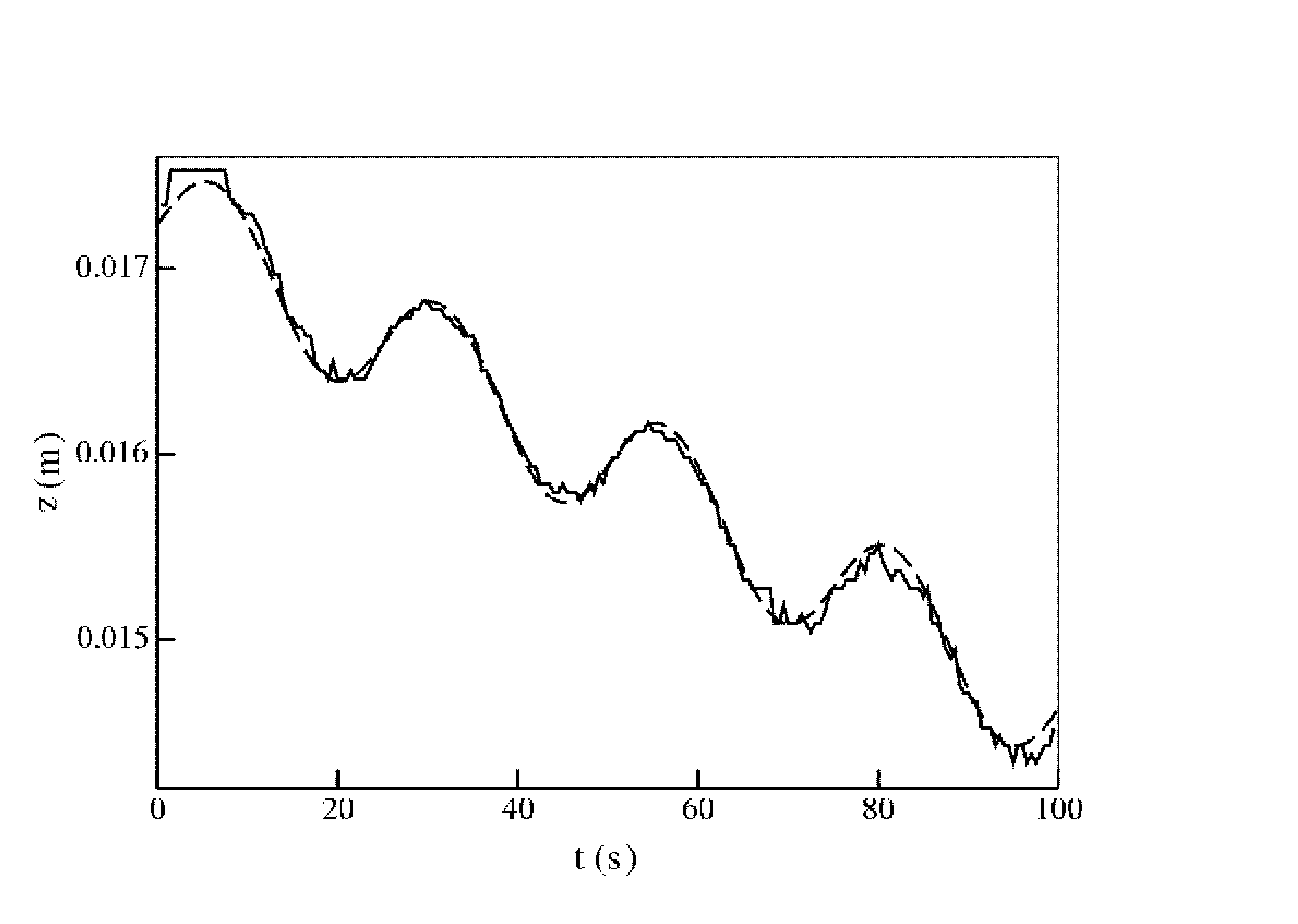,width=70mm,angle=-90}
    \end{minipage}
  \end{center}
  \caption{\small Time evolution of the front position.
   $-$: Experimental data, $--$: Numerical simulations.
    The experimental values
    $(A,f)$ ($A$ in $mm$ and $f$ in $Hz$) are: $(1.12,0.01)$, $(0.55,0.01)$,
    $(0.14,0.08)$, $(0.27,0.04)$, $(0.1,0.06)$,
    $(0.14,0.04)$ from top left and clockwise.}
  \label{superpose_exp_nu}
\end{figure}

It is seen from these movies that the front oscillates and travels
from the burnt product to the fresh reactant. The mean
concentration profiles are obtained by averaging along the lattice
width, and analyzed along the same line as in the experiments.
Figure \ref{superpose_exp_nu} shows the time evolution of the
displacement of the iso-concentration $\overline{C}=0.5$, obtained
in the $2D$ simulations and in the experiments: The agreement
between the two supports the contention that in our frequency
range, the dynamics of the front is governed only by the
variations of the velocity field in the gap ($b=0.4\; mm$), and
that the (large) transverse extent of the plane of the
experimental cells ($h=8\; mm$) plays no role.

\begin{figure}[!h]
  \begin{center}
    \begin{minipage}{110mm}
      \psfig{file=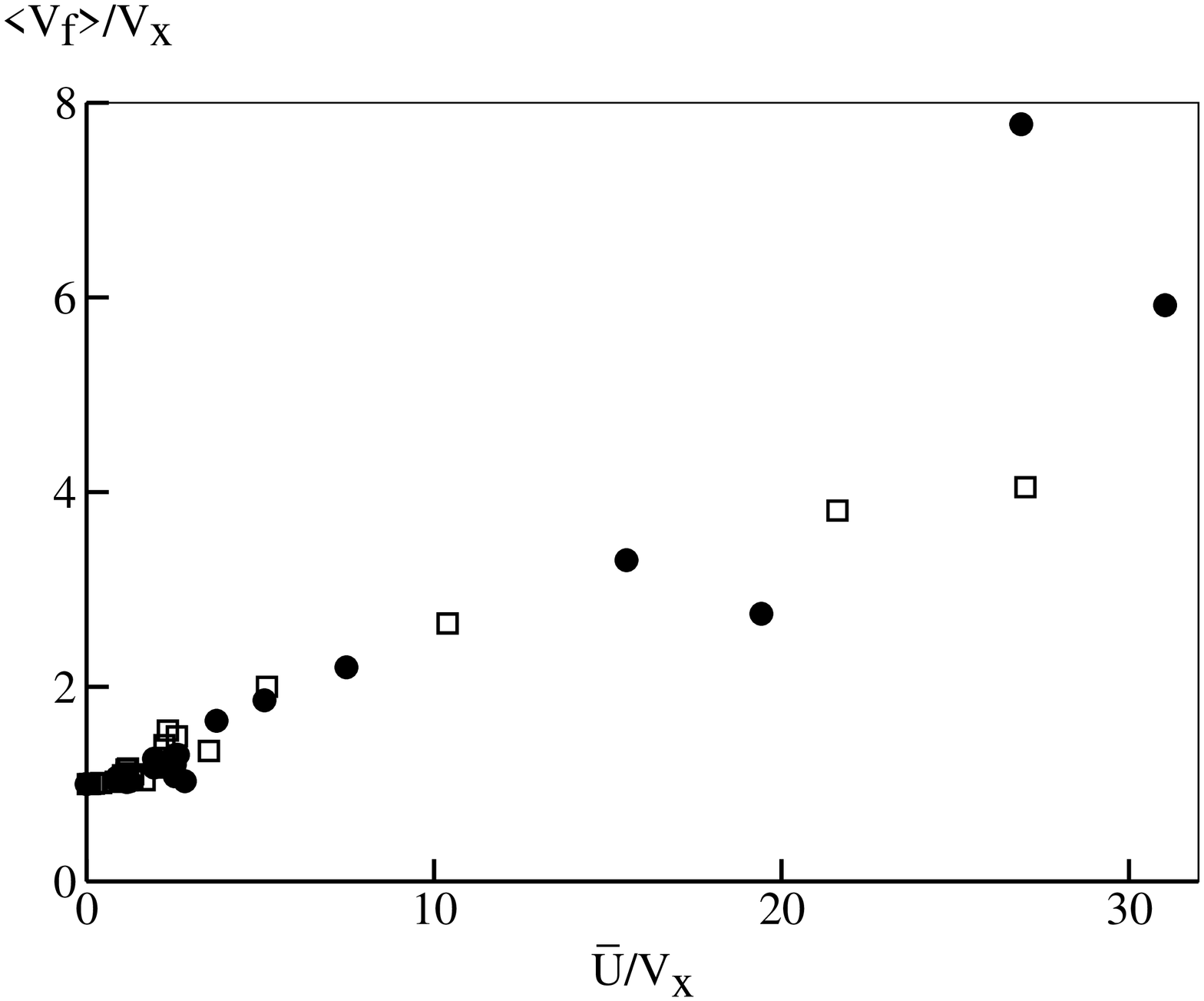,width=110mm,angle=-90}
    \end{minipage}
  \end{center}
  \caption{\small Normalized drift velocity
    $\langle V_f\rangle /V_\chi$ versus the normalized flow
    velocity $\overline{U}/V_\chi$.
    $\bullet$: Experimental data, {\tiny$\square$}: Numerical
    simulations.}
  \label{vit_moy_num}
\end{figure}

Figure \ref{vit_moy_num} displays the resulting normalized drift
front velocity $\langle V_f\rangle/V_\chi$ versus the normalized
flow velocity $\overline{U}/V_\chi$. The two sets of data,
obtained experimentally and numerically, are also in good
agreement. This leads us to analyze in more details the dynamics
of the front in the gap, with the help of the $2D$ numerical
simulations.

The theoretical front width given in (\ref{co}) can be obtained
from the second-order moment of the derivative of the
concentration profile \cite{leconte04}:
\begin{equation}
  l_\chi^2 = \frac{3}{\pi^2}
  \int_{-\infty}^{+\infty}z^2\frac{d\overline{C}}{dz}dz
  \label{frwidth}
\end{equation}

\begin{figure}[!h]
  \begin{center}
    \begin{minipage}{130mm}
      \psfig{file=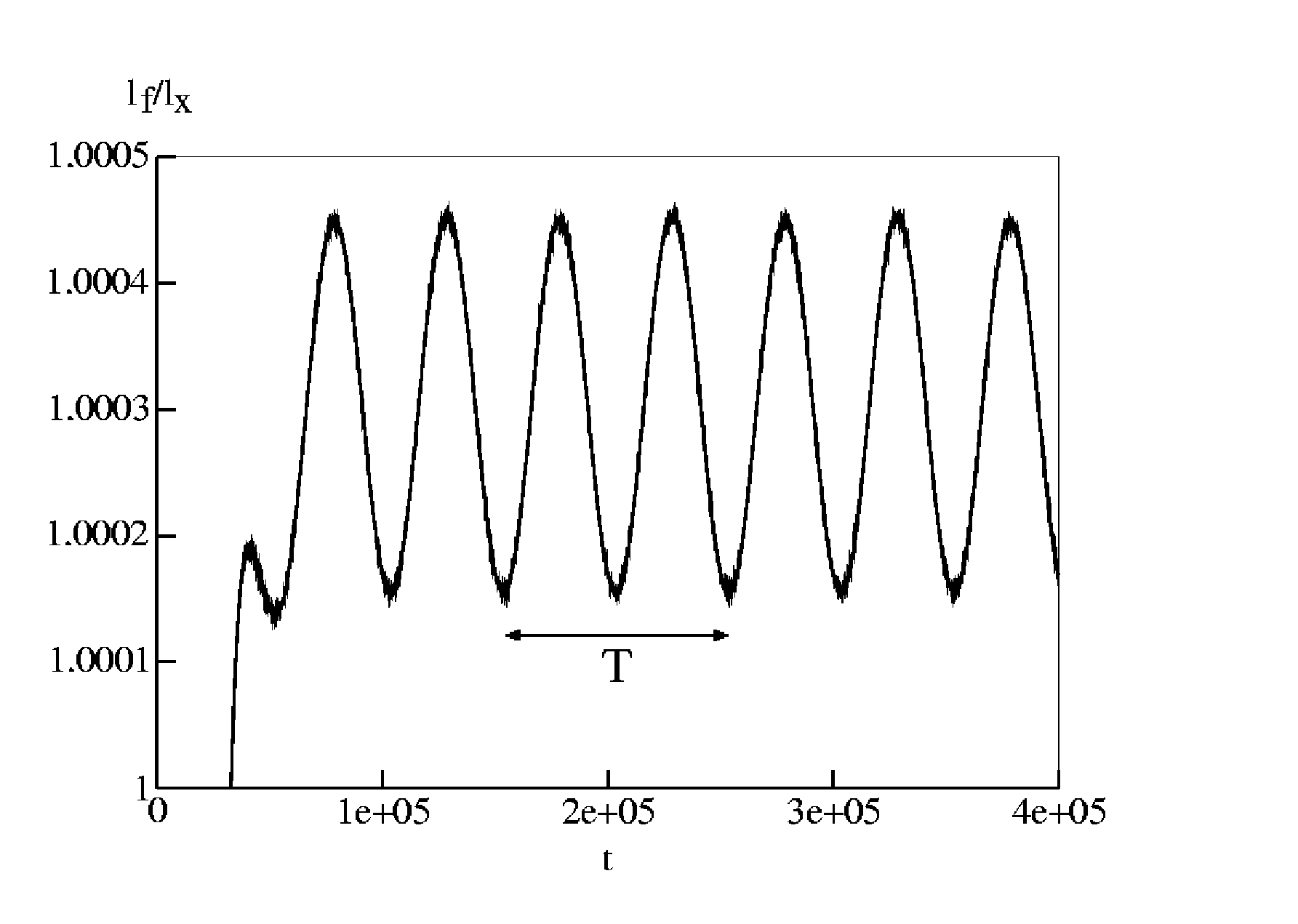,width=130mm,angle=-90}
    \end{minipage}
  \end{center}
  \caption{\small Time evolution of the normalized effective front width
    $l_f/l_\chi$ for the numerical simulation $A=10$ and $f=10^{-5}$. The period $T$ of the
     oscillating flow is indicated for comparison.}
  \label{largeur}
\end{figure}

The use of the discrete version of (\ref{frwidth}) allows us to
estimate the front width $l_f$. Figure \ref{largeur} displays the
time evolution of $l_f$.
After a transient time of the order of one flow period, the front
width $l_f(t)$ becomes periodic and oscillates with a frequency
twice that of the imposed flow, in agreement with the
experimental result given in figure \ref{mic_dop}.\\

\begin{figure}[!h]
  \begin{center}
    \begin{minipage}{130mm}
      \psfig{file=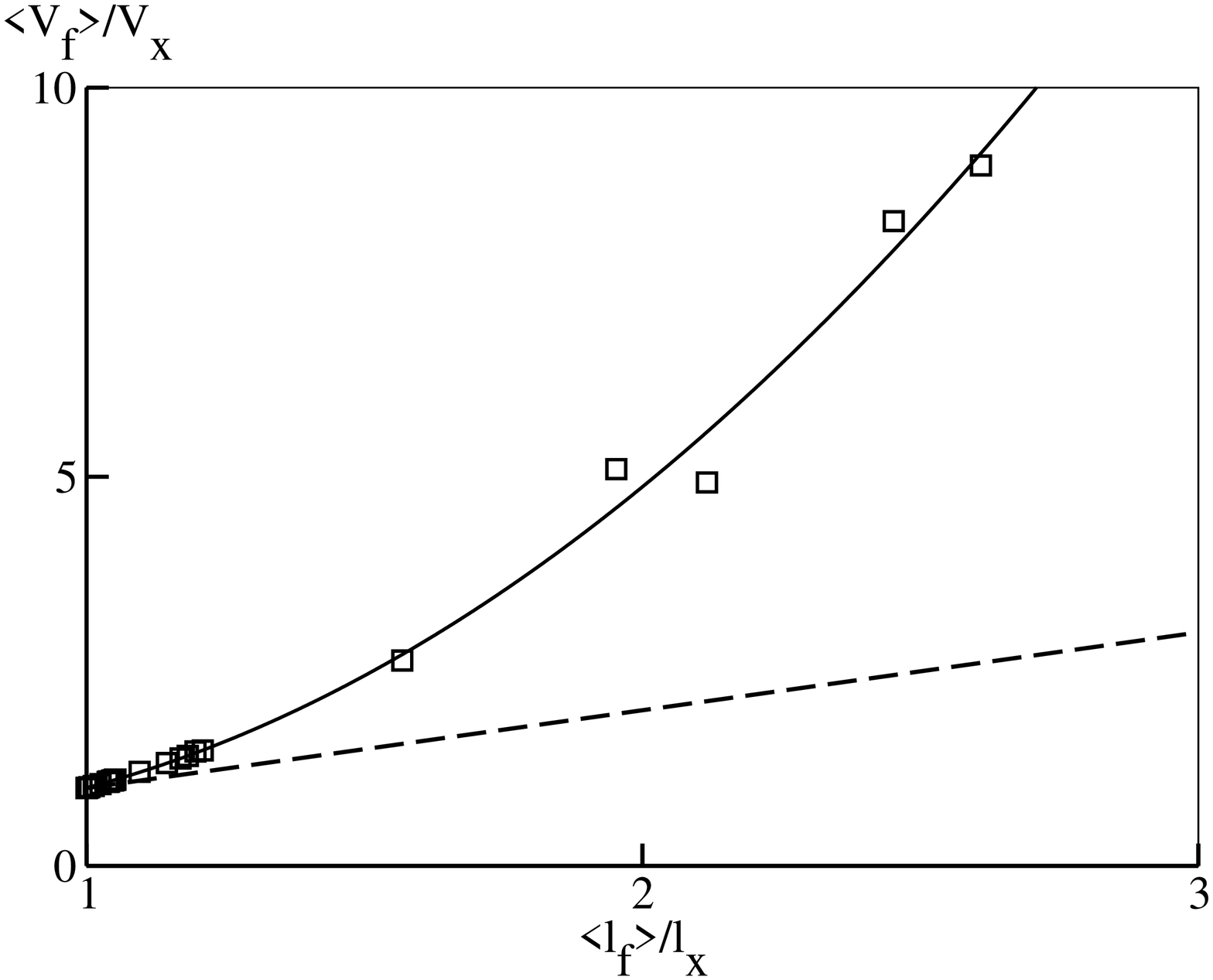,width=130mm,angle=-90}
    \end{minipage}
  \end{center}
  \caption{\small {Normalized front velocity
    $\langle V_f\rangle /V_\chi$ versus normalized front width
    $\langle l_f\rangle /l_\chi$. The line through the data (best
    quadratic fit) indicates that $\langle V_f\rangle /V_\chi$ is not
     proportional to $\langle l_f\rangle /l_\chi$. Dashed line :
     $\langle V_f\rangle /V_\chi = \langle l_f\rangle /l_\chi$.}}
  \label{lv}
\end{figure}

For a stationary laminar flow, we showed in \cite{leconte04} that,
under mixing regime conditions, the velocity and width of the
reaction front result from a Taylor-like diffusion process, with
an effective diffusion coefficient $D_{eff}$ such that
$V_f=\sqrt{\alpha D_{eff}/2}$ and $l_f=\sqrt{2D_{eff}/\alpha}$,
leading to $ V_f / l_f = \alpha /2$. An easy way to test the
relevance of this effective diffusion description to the present
case is to measure the ratio of the time-averaged values $\langle
V_f\rangle/\langle l_f\rangle$. Figure \ref{lv} clearly shows that
the relation between $\langle V_f\rangle$ and $\langle l_f\rangle$
is not linear. This allows us to discard the above description.

The information given by both the experiments and the numerical
simulations may be summarized as follows: In the presence of a
periodic flow, the propagation of a chemical front in a Hele-Shaw
cell is governed by the velocity profile in the gap. The front
position drifts in the natural direction of the chemical wave and
undergoes oscillations at the frequency of the flow whereas the
front width oscillates at twice this frequency. Finally, the front
behavior cannot be described in terms of a Taylor-like effective
diffusion.

In the next section, we derive a model, under the assumption of
weak flow velocity, and discuss the above findings in the light of
our
theoretical approach.\\

\section*{Theoretical determination of the front velocity}

We have derived in a previous work \cite{leconte04}, the velocity
of a stationary reaction front, using a small parameter expansion
method. The two parameters were the reduced gap thickness
$b/l_{\chi}$ (or $\eta=b/2l_\chi$) and the normalized flow
velocity $\varepsilon=\overline{U}/V_{\chi}$, with
$\overline{U}=2U_{M}/3$ in $2D$. We shall use the same method in
order to derive the instantaneous front velocity and width. We
note that this extends the work by Nolen and Xin \cite{nolen03} on
the drift front velocity.

The ADR equation in a $2D$, unidirectional flow along $z$ writes:
\begin{eqnarray}
  \frac{\partial C}{\partial t}+U\frac{\partial C}{\partial z} =
  D_m\left (\frac{\partial^2 C}{\partial x^2} +
  \frac{\partial^2 C}{\partial z^2} \right ) + \alpha C^2(1-C)
  \label{adre}
\end{eqnarray}
The $2D$ flow is imposed in a gap of size extension $b$. We assume
a frequency small enough ($l_{\nu}=\sqrt{\nu/\omega}\gg b$) to
have an oscillating Poiseuille flow:
\begin{equation}
  U(x,t)=U_M\left ( 1-\frac{4x^2}{b^2}\right )\sin{(\omega t)}
  \label{poise}
\end{equation}

We also assume that the flow velocity is small compared to
$V_\chi$,
\begin{equation}
  \varepsilon = |U_M|/V_\chi \ll 1 \label{hyp1}
\end{equation}
and that, accordingly,  the change in the front velocity due to
the presence of the flow is small compared to $V_\chi$,
$(V_f(t)-V_\chi) \ll V_\chi$.

Moreover, we assume that the concentration field is nearly uniform
along the transverse $x$ direction. Note that, for a passive
tracer, this hypothesis is fulfilled when the P\'eclet number
$Pe=|U|b/D_m$ is smaller than $L/b$, where $L$ is the typical
advection length (condition for Taylor-diffusion regime). In the
presence of a reaction, this condition becomes $\varepsilon\eta^2
\ll 1$ \cite{leconte04}. We recall first the approach of Nolen and
Xin \cite{nolen03}, in our notations. In the moving frame
($s=z-V_f t$), under the assumption (\ref{hyp1}) of a weak flow
velocity ($\varepsilon \ll 1$), the concentration $C(s,x,t)$ and
the velocities $U$ and $V_f$ can be expanded in powers of
$\varepsilon$, as follows:
\begin{eqnarray}
  C&=&C_0(s) + \varepsilon C_1(s,x,t) \label{c}\\
  U&=& \varepsilon U_1(x,t) \label{u}\\
  V_f&=&V_\chi + \varepsilon V_1(t) + \varepsilon^{2} V_2(t) + ... \label{v}
\end{eqnarray}
where $C_0(s)=1/(1+e^{s/l_f})$ is the mean concentration profile
(averaged over the gap and the time) and $C_1(s,x,t)$ denotes
deviations from the mean. Using the space and time Fourier
decomposition of the flow velocity field,
$U_{1}=\sum_{n}b_{n}e^{ik_{n}x+i\omega t}$, for a monochromatic
($\omega$) velocity field (where $k_{n}=2n\pi /b$ is the
decomposition wave vector for a gap of width $b$), and expanding
the ADR equation in the moving frame,  Nolen and Xin derived the
time-averaged, drift velocity:
\begin{equation}
  \langle V_{f} \rangle=V_{\chi}(1+\varepsilon^{2}\gamma/2) \mbox{, with }
  \gamma=\frac{\eta^2}{2\pi^2}
  \sum_{n>0,}\frac{2n^{2}\mid b_{n}\mid^{2}}{
    \left (n^{4}+\frac{\omega^{2}}{\Omega^{2}}\right )}
  \label{nolxin}
\end{equation}
where $\Omega =\frac{4\pi^2D_m}{b^2}$ is a characteristic
frequency, proportional to the inverse of the typical diffusion
time across the gap $b$. For the Poiseuille flow used here, one
finds $b_{n}=(-1)^{n}3/\pi^{2}n^{2}$.

In this paper, we are interested in the time dependence of the
front velocity, in the presence of a sine flow velocity  (a single
temporal mode, $\omega$, in the Fourier decomposition of $U_{1}$).
After some calculations and with the necessary assumption
$\frac{\partial^2 C_1}{\partial x^2} \ll \frac{\partial^2
C_1}{\partial s^2}$, we find (see \cite{marcthese} for details):

\begin{equation}
  \frac{V_f}{V_\chi}(t)=1 +\varepsilon \cos{(\omega t)}+\varepsilon^{2}\eta^{2}
  [\gamma_1 (1+ \cos{(2\omega t)})+ \gamma_2 \sin{(2\omega t)}]
  \label{v2vx}
\end{equation}
with
\begin{eqnarray}
  \gamma_1&=&\frac{9}{2\pi^6} \sum_{n> 0}
  \frac{1}{n^2\left ( \frac{\omega^2}{\Omega^2}+ n^4\right )}\label{v2vxbis}\\
  \gamma_2&=&\frac{9}{2\pi^6}\frac{\omega}{\Omega}
  \sum_{n> 0}\frac{1}{n^4\left ( \frac{\omega^2}{\Omega^2}+ n^4\right )}\label{v2vxter}
\end{eqnarray}

\begin{figure}[!h]
  \begin{center}
    \begin{minipage}{170mm}
      \hspace{-6mm}\psfig{file=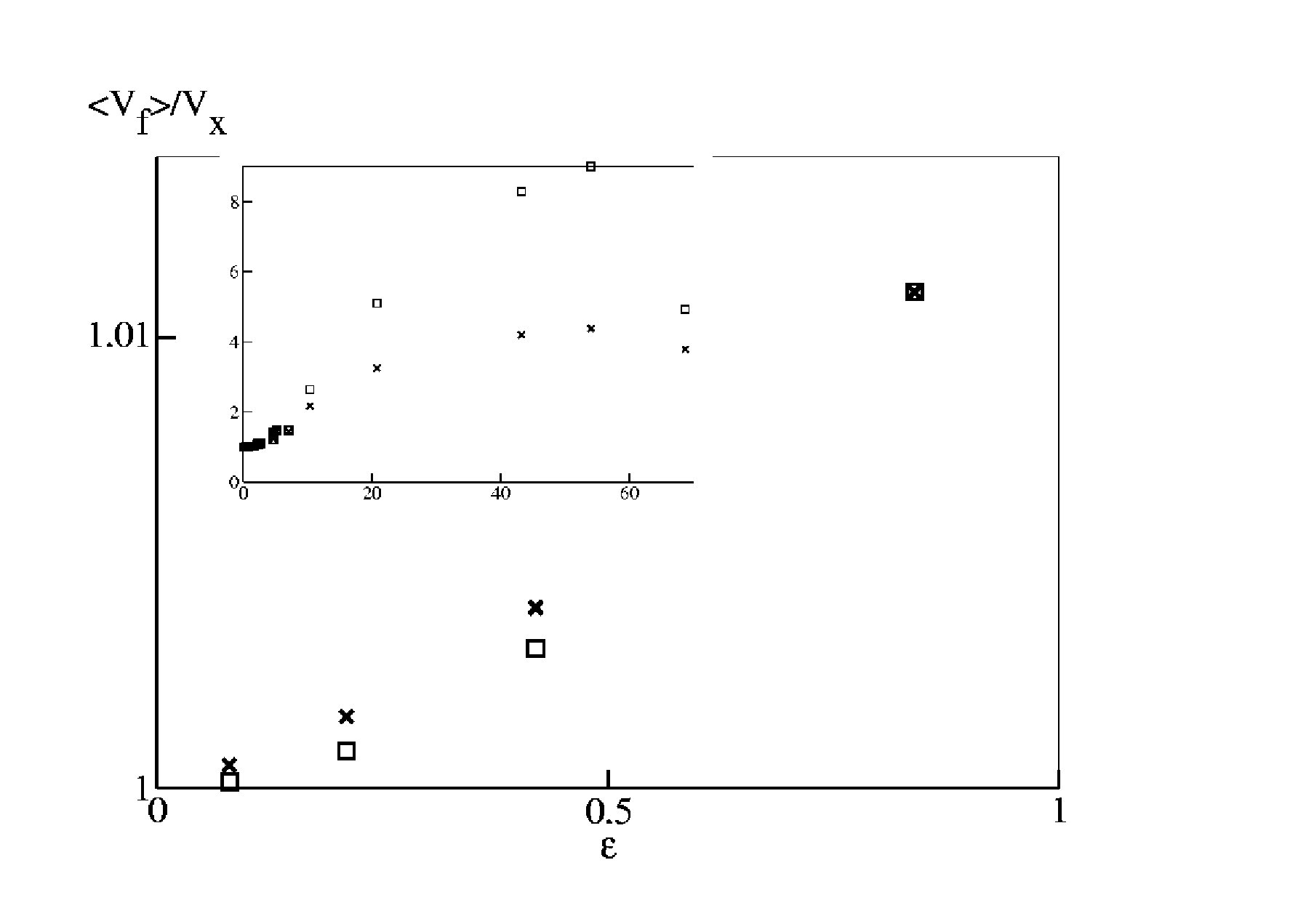,width=90mm,angle=-90}
      \hspace{-5mm}\psfig{file=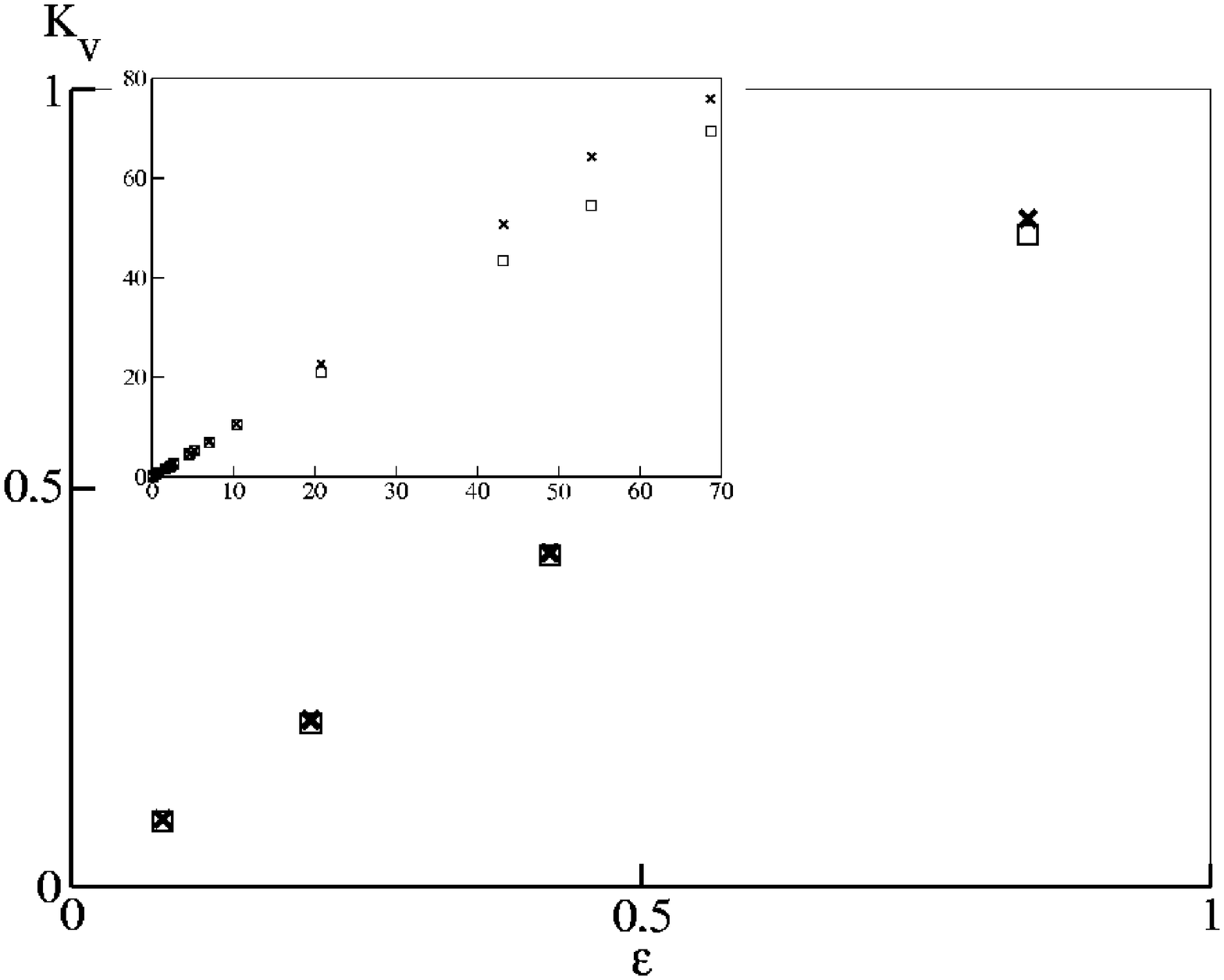,width=90mm,angle=-90}
    \end{minipage}
  \end{center}
  \caption{\small Left: Normalized drift front velocity
    $\langle V_f\rangle /V_\chi$ versus $\varepsilon = |U_M|/V_\chi$.
    {\tiny$\square$}: Numerical simulation data,
    $X$: Theoretical predictions in the limit of small $\varepsilon$.\newline
    Right: Amplitude $K_{v}$ of the oscillation of $V_f(t)/V_\chi$
    versus $\varepsilon$. {\tiny$\square$}:
    Simulations results, $X$: Theoretical results.\newline
    The insets give the behavior of $\langle V_f\rangle/V_\chi$ and $K_v$
    over a wider range of $\varepsilon$.}
  \label{comp_num_theo2}
\end{figure}

The equations (\ref{v2vx}), (\ref{v2vxbis}) and (\ref{v2vxter})
display no term involving $\omega/\alpha$, which demonstrates that
the reaction kinetics does not play any role at this order. At
first order in $\varepsilon$ the front velocity is the algebraic
sum of the gap-averaged flow velocity and of the chemical wave
velocity in the absence of flow. Note that this result is
surprisingly similar to the one obtained for the front
displacement in constant flows in the mixing regime
\cite{edwards02,leconte04}. The leading order pulsative flow
contribution causes the front to oscillate with the flow
frequency. At second order in $\varepsilon$, the contributions to
the front velocity are a constant one and one at twice the
frequency of the flow field. Note that for $\omega=0$, we recover
previous results: $\frac{V_f}{V_\chi}\approx
1+\varepsilon+\frac{1}{105}\varepsilon^2\eta^2$
\cite{papanicolaou91,edwards02,leconte04}.
Averaging expression (\ref{v2vx}) over one period leads to the
drift front velocity (\ref{nolxin}) $\langle V_{f}^{theo}\rangle
/V_{\chi}=1+\varepsilon^{2}\eta^{2} \gamma_{1}$. We note that the
amplitude $K_{v}$ of the velocity oscillation is mainly dominated
by the order one term: $K_{v}\simeq\varepsilon$. Figure
\ref{comp_num_theo2} shows that the theoretical results and the
numerical simulation data for the drift velocity and its amplitude
of oscillation are in reasonable agreement for small $\varepsilon$
($\varepsilon<1$).

\section*{Conclusion}

In this paper, we have analyzed the influence of a time-periodic
flow on the behavior of an autocatalytic reaction front. The
numerical simulations are in reasonable agreement with the
experiments and show that the front dynamics is controlled by the
flow velocity in the gap of a Hele-Shaw cell in the range of the
parameters explored. We have shown that, contrary to the case of a
stationary laminar flow, a Taylor-like approach cannot account for
the (time-averaged) width and velocity of the front. The
instantaneous front velocity has been derived theoretically, in
the weak flow velocity regime. It is found to be in reasonable
agreement with the numerical simulation results for $|U_M|/V_\chi
< 1$.

\section*{Acknowledgments}

We thank Dr C\'eline L\'evy-Leduc for fruitful discussions. This
work was partly supported by IDRIS (Project No. 034052), CNES (No.
793/CNES/00/8368), ESA (No. AO-99-083) and a MRT grant ( for M.
Leconte). All these sources of support are gratefully
acknowledged.

\section*{Appendix : Determination of the oscillating velocity
profile in a Hele-Shaw cell}

We consider a laminar flow, unidirectional along the $z$
direction, in a Hele-Shaw cell of cross-section $b\times h$ ($b
\ll h$) in the $x$ and $y$ directions, respectively. The
Navier-Stokes equation in the $z$ direction writes:
\begin{equation}
\frac{\partial U}{\partial t} = \nu \left (\frac{\partial^2
U}{\partial x^2}+\frac{\partial^2 U}{\partial y^2} \right ) -
\frac{1}{\rho}\frac{\partial P}{\partial z}e^{-i\omega t}
 \label{ns}
\end{equation}
where a harmonic pressure gradient oscillating at frequency
$f=\omega/2\pi$ is imposed, and the flow velocity $U(x,y,t)$
satisfies the boundary conditions:
\begin{equation}
U(x,\pm \frac{h}{2},t) = U(\pm \frac{b}{2},y,t) = 0 \label{cl}
\end{equation}

In order to derive the analytical expression of $U$, we extend the
method used by Gondret {\it et al} \cite{gondret97} to calculate
the stationary flow field. First, we consider the flow field,
written $U^*(x)e^{-i\omega t}$, between two infinite planes (at
$x=\pm \frac{b}{2}$ as in our simulations), which satisfies:
\begin{equation}
\frac{\partial^2 U^*}{\partial x^2} + \frac{i\omega}{\nu} U^* =
\frac{1}{\rho \nu}\frac{\partial P}{\partial z} \label{eq1}
\end{equation}
leading to \cite{landau89}:
\begin{equation}
U^*(x) = \frac{1}{i\omega \rho}\frac{\partial P}{\partial z} \left
[ 1 - \frac{\cos{(kx)}}{\cos{\left (\frac{kb}{2}\right )}} \right
] \label{soleq1_bis}
\end{equation}
where
$k=\sqrt{\frac{i\omega}{\nu}}=(1+i)\sqrt{\frac{\omega}{2\nu}}$.
Following \cite{gondret97}, we write now the solution of the full
problem as:
\begin{equation}
U(x,y,t)= \left [ U^*(x) + U^{**}(x,y)\right ] e^{-i\omega t}
\label{u}
\end{equation}
where $U^{**}(x,y)$ must now satisfy:
\begin{equation}
\frac{\partial^2 U^{**}}{\partial x^2} + \frac{\partial^2
U^{**}}{\partial y^2} + \frac{i\omega}{\nu}U^{**} = 0 \label{eq2}
\end{equation}
with the boundary conditions:
\begin{eqnarray}
U^{**}(\pm \frac{b}{2},y) &=& 0
\label{cond1} \\
U^{**}(x,\pm \frac{h}{2}) &=& -U^{*}(x) \label{cond2}
\end{eqnarray}

The solution of (\ref{eq2}) is found assuming a Fourier
decomposition of the form:
\begin{equation}
U^{**}(x,y)=\sum_{n=1}^{\infty}A_n \cos{(k_{xn}x)}\cos{(k_{yn}y)}
\label{solgeneq2}
\end{equation}
where $k_{xn} = (2n-1)\pi/b$, in order to satisfy the boundary
condition (\ref{cond2}). After some calculations, we obtain the
solution of (\ref{ns}):
\begin{equation}
U(x,y,t)=-\frac{ie^{-i\omega t}}{\omega \rho}\frac{\partial
P}{\partial z} \left [ 1 - \frac{\cos{(kx)}}{\cos{\left
(\frac{kb}{2}\right )}} + \sum_{n=1}^{\infty}4(-1)^n
\frac{k^2\cos{(k_{xn}x)} \cos{(k_{yn}y)}}{(2n-1)\pi
k_{yn}^2\cos{\left (k_{yn}\frac{h}{2}\right )}} \right ]
\label{soleq2}
\end{equation}
where $k_{yn}^2=\frac{i\omega}{\nu}-k_{xn}^2$. The velocity
$U(x,y,t)$  induced by an oscillating pressure gradient has a
phase shift and a modulus which depend on space. We note that the
maximum modulus $U_{M}$ occurs in the middle of the cell
($x=y=0$). In Expression (\ref{soleq2}), the first two terms
correspond to a 2D oscillating flow between two parallel
boundaries at distance $b$ ($U^{*}$) \cite{landau89}, and the
third one accounts for the finite size of the cell in the larger
direction $h$ ($U^{**}$). We note that at high frequency, a plug
flow takes place in the section of the cell, except in a thin
viscous layer $l_{\nu}=\sqrt{\nu/\omega}$ close to the boundaries
($y=\pm h/2$
 and  $x=\pm b/2$). On the opposite, in the low frequency regime
($l_{\nu}\gg b$), the oscillating flow has the same shape as a
static one given in \cite{gondret97}. Hence, except in a thin
layer (of thickness  $\sim b \ll h$) close to the side boundaries
($y=\pm h/2$), the flow is a parabolic Poiseuille flow of the
form:
\begin{equation}
  U(x,t)=U_M\left ( 1-\frac{4x^2}{b^2}\right )\sin{(\omega t)}
\end{equation}



\end{document}